\documentclass[fleqn,10pt,lineno]{wlscirep}

\usepackage{tabularx}
\newcolumntype{L}{>{\raggedright\arraybackslash}X}
\usepackage{textcomp}
\usepackage{url}
\providecommand{\doi}[1]{doi:\,#1}

\title{Horizon-resolved Forecastability of Time Series via Auto Mutual Information}

\author[1]{Peter M. Catt\thanks{Corresponding author: peter.catt@virtualblue.co.nz}}
\affil[1]{Virtual Blue Ltd, Auckland, New Zealand}


\begin{abstract}
In many social, business, economic, and physical systems the true data-generating process is rarely known, so a series' forecastability cannot be assumed; it must be assessed from the observed history. This paper evaluates a horizon-specific, pre-modelling measure for that assessment: auto-mutual information (AMI), a training-only measure of past-future dependence at each forecast horizon; higher AMI indicates more recoverable temporal structure. It is evaluated on the M4 Monthly dataset (47,992 of 48,000 series, 18-month horizon) with seasonal naïve, ETS, and N-BEATS as probes and MASE as the error measure. The series-level association with realised skill is modest but systematic (mean Spearman $\rho$ of 0.11 to 0.13 for ETS and N-BEATS), accumulating into a pronounced descriptive gradient: median MASE is 29\% (ETS) to 37\% (N-BEATS) lower in the top within-horizon AMI decile than in the bottom. AMI outperforms absolute autocorrelation as a horizon-specific diagnostic by a small, consistent margin; paired bootstrap intervals exclude zero for all three probes, and synthetic benchmarks with analytically exact AMI locate the advantage on nonlinear dependence. Decile assignments are sample-relative and require recalibration on a new portfolio. The contribution is a pre-modelling diagnostic, not a forecasting model.
\end{abstract}

\begin{document}

\flushbottom
\maketitle
\thispagestyle{empty}

\makeatletter
\providecommand{\@keywords}{}
\ifx\@keywords\@empty\else
  \noindent\textbf{Keywords:} \@keywords\par\medskip
\fi
\makeatother

Forecast accuracy becomes observable only after a model has been fitted and its errors realised. How much a series' own history constrains its future is, by contrast, a property of the observed record, and can be measured before any model is fitted. The shift from single-series to large-portfolio forecasting is well documented in both the academic and practitioner literature \cite{syntetos2016,boylan2021,petropoulos2022}. Such portfolios are heterogeneous: some series respond well to flexible nonlinear models and others do not, and average-case benchmarking obscures this variation \cite{moniz2017}. Large-scale evidence from the M4 competition shows that simple statistical methods are on average highly competitive with sophisticated machine-learning approaches \cite{makridakis2020}. Many operational systems nevertheless apply a single modelling pipeline uniformly across all series, categories, and horizons, and automated machine-learning pipelines reduce the manual effort without measuring the temporal information each series contains \cite{wangmc2026}.

Most forecasting evaluation is retrospective: a model is trained, forecasts are generated, and accuracy is observed after the fact. Forecasting support systems are organised around this accuracy-observation loop \cite{fildes2006}, and even principled selection rules derived from retrospective accuracy generalise poorly across portfolios \cite{fildes2015}, so substantial modelling effort is invested before its payoff is known. What the loop does not answer is a prior, pre-modelling question: does a series' history contain exploitable temporal structure at the horizons that matter, and how much? A diagnostic that answered this before model training would let forecastability be assessed, compared across series and horizons, and tracked over time without first fitting models. Measuring that quantity, before any model is fitted, is the problem this paper addresses.

Several families of pre-modelling diagnostic exist. Autocorrelation and partial autocorrelation provide a linear summary of temporal dependence at each lag, and seasonality strength and trend strength summarise structural properties of a series \cite{hyndman2008}; these are useful but collapse dependence into linear or structured summaries and may miss nonlinear temporal patterns that flexible machine-learning models can exploit. Entropy-based measures, including approximate, sample, and permutation entropy \cite{pincus1991,richman2000,bandt2002}, have been proposed as complexity measures associated with forecast difficulty \cite{catt2009,pennekamp2019,ponceflores2020,papacharalampous2022,garland2014}, and forecastable component analysis defines forecastability via spectral entropy \cite{goerg2013}. Feature-based meta-learning approaches \cite{lemke2010,monteromanso2020,talagala2023,santos2025,prudencio2004,kang2017,bahrpeyma2025,li2026} map collections of series features to recommended model classes, but require large libraries of historical accuracy data and offer no direct, horizon-specific signal about whether exploitable information exists. More recent work by Wang et al. \cite{wang2025} proposes scalar forecastability measures, a spectral predictability score and a Lyapunov-exponent-based measure, evaluated against realised forecast performance; both compress a series to a single number with no horizon resolution, and model selection in large portfolios nevertheless remains largely heuristic or computationally brute-force.

This paper proposes auto-mutual information (AMI) as an interpretable, model-agnostic, horizon-specific pre-modelling measure of time-series forecastability. AMI measures statistical dependence between present and future values of a series at each forecast horizon: in plain operational terms, higher AMI means more recoverable temporal structure at that horizon, so more of the future is encoded in the present observation and a model has more to work with. The core framing is direct: AMI is not a forecasting model but a diagnostic of where a series retains exploitable past-future dependence at the horizon of interest. It occupies a distinct position among the diagnostic families above. Unlike autocorrelation, it captures nonlinear dependence. Unlike entropy measures, it is horizon-specific and directly interpretable as a measure of past-future information. Unlike meta-learning, it requires no external training data, measures temporal dependence directly instead of through proxy features, and produces a signal grounded in information theory rather than empirical accuracy libraries. That temporal dependence and forecastability are related is theoretically expected and well understood \cite{shannon1948,bialek2001,delsole2004,delsole2005,delsole2007,coverthomas2006}; the contribution is an operationalisation and large-scale evaluation of horizon-specific forecastability measurement. The prior entropy-based work cited above established the concept at small scale. What is new here is that AMI is quantified horizon-specifically, estimated non-parametrically from a fixed training window, and evaluated at operational scale on a canonical benchmark, producing category-resolved forecastability profiles before any model is fitted, with a more precise estimator and a model panel spanning simple seasonal benchmarks to flexible neural forecasting.

For a stationary time series $Y_t$, the horizon-specific auto-mutual information is defined as:

\begin{equation}\label{eq:ami}
\mathrm{AMI}(h) = I\!\left(Y_t;\, Y_{t+h}\right)
\end{equation}

where $I(\cdot;\cdot)$ denotes the mutual information between $Y_t$ and $Y_{t+h}$, computed from their joint distribution relative to the product of their marginals. Shannon \cite{shannon1948} defined mutual information as the reduction in uncertainty about one variable given knowledge of another; Bialek et al. \cite{bialek2001} formalised this as predictive information in the context of learning and forecasting. The definition is stated for a stationary process. On an observed, possibly non-stationary series the quantity estimated is the empirical past-future dependence of the series over the training window, not the AMI of an underlying stationary process; this estimand is made precise, and the choice defended, in the Methods.

Several properties make AMI directly relevant as a forecasting diagnostic. AMI equals zero if and only if $Y_t$ and $Y_{t+h}$ are statistically independent, and it is non-negative \cite{coverthomas2006}. It captures nonlinear dependence that autocorrelation misses \cite{li1990}, is horizon-specific, and can be estimated from training data before any model is fitted. Jaynes \cite{jaynes1957} and Grassberger \cite{grassberger1986} provide the deeper information-theoretic foundations; Crutchfield and Feldman \cite{crutchfield2003} connect predictive information to excess entropy and statistical complexity.

AMI is a long-established quantity in nonlinear time-series analysis, where Fraser and Swinney \cite{fraser1986} introduced it to select an embedding delay for phase-space reconstruction, a practice since standardised in the field \cite{kantz2004}. That procedure consumes a single point of the curve, the first local minimum, and the lag it returns is selected by a criterion orthogonal to any forecasting decision: it is chosen because dependence there has stopped falling, which makes successive reconstruction coordinates informative about one another without being redundant. The remainder of the curve is not interpreted. The present use inverts that reading. AMI is evaluated at the horizon the forecasting decision fixes, and the quantity of interest is the dependence retained at that horizon rather than the lag at which dependence is locally weakest.

A central conceptual distinction in this paper is between forecastability and model competence. Forecastability refers to the information available in the declared information set: how much that set reduces uncertainty about the future at a given horizon. Model competence refers to how well a specific model exploits that information. These are logically independent, and the distinction matters when interpreting realised error: low forecast error can arise because a series is genuinely forecastable or merely because the task is easy through scale, smoothness, or strong seasonality, whilst high forecast error can reflect little predictive information or a model that fails to exploit the information available. AMI addresses the first stage, measuring whether exploitable temporal information appears to exist before model fitting begins: it asks whether exploitable past-future dependence exists before asking which model should exploit it.

Because AMI is computed entirely from the training data prior to any model fitting, it is a strictly pre-modelling signal that can be generated automatically across an entire portfolio as a feature-engineering layer. Two scope constraints apply throughout. First, the decile assignments used for stratified description are sample-relative, computed within the evaluation panel at each horizon; they are not a universal forecastability scale and are not directly portable to an arbitrary new portfolio without recalibration. Second, the diagnostic is computed on a declared information set consisting of the single lagged observation $Y_{i,t}$ at each horizon, so it sees neither the remainder of the series history nor information carried by external drivers, hierarchical relationships, calendar effects, or other series in the portfolio.

The paper evaluates whether horizon-specific AMI, estimated from training data alone, ranks series-horizon pairs by realised forecast error. It makes three contributions: it operationalises AMI as an automated, horizon-specific, non-parametric pre-modelling measure of forecastability and evaluates it at scale on the M4 Monthly benchmark treated as a heterogeneous portfolio; it provides a direct horizon-wise comparison against absolute autocorrelation, with paired bootstrap uncertainty on the difference; and it validates the estimator on synthetic benchmarks with analytically exact AMI, locating that advantage on nonlinearly transformed processes.

\section*{Results}

\subsection*{AMI reveals heterogeneous forecastability}

\begin{figure}[tbp]
\centering
\includegraphics[width=\textwidth]{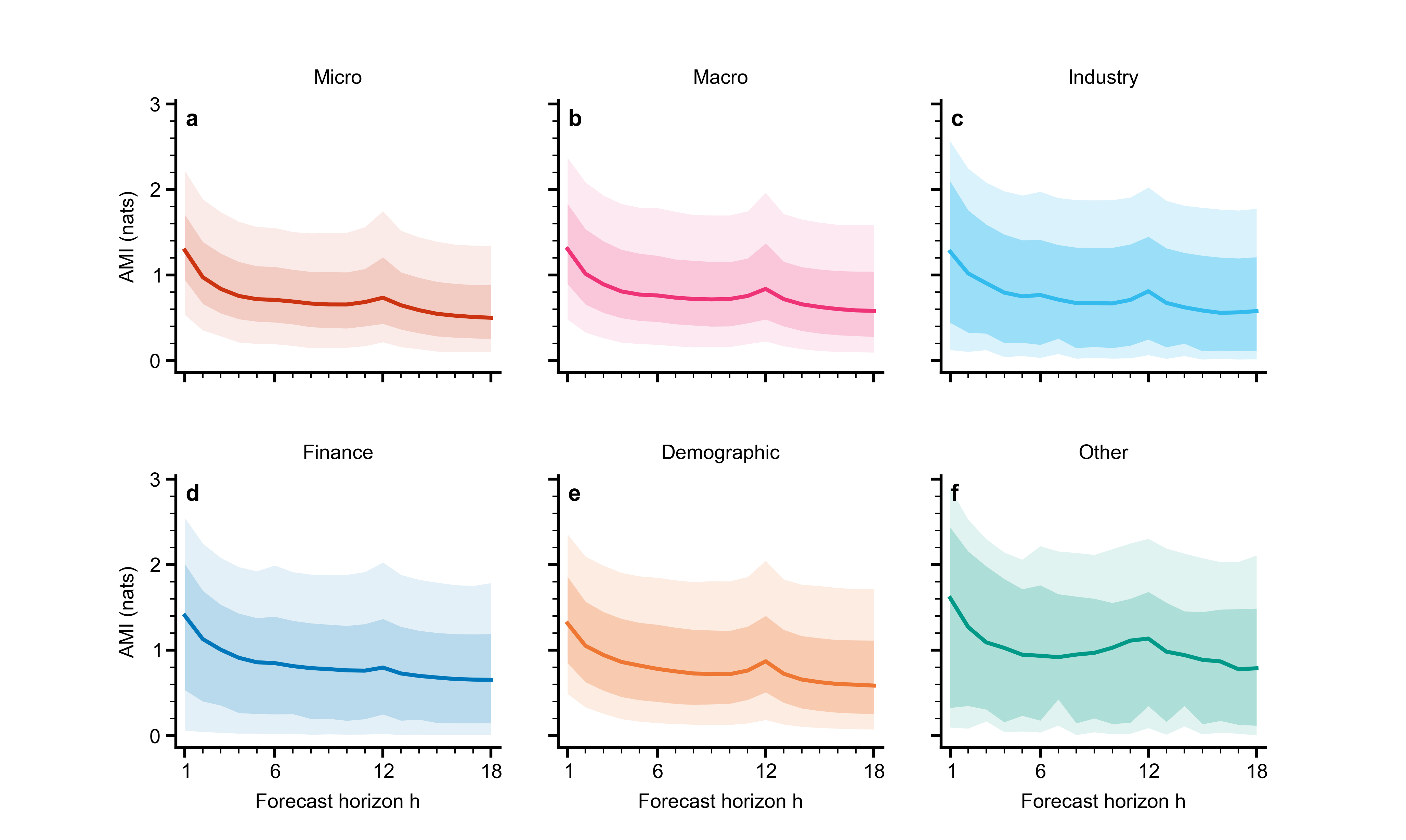}
\caption{AMI profiles by M4 Monthly category. One panel per category: the line is the median AMI (nats) across that category's series at each forecast horizon h = 1 to 18, with the interquartile range (darker band) and the 10th-90th percentile range (lighter band), all estimated from training data only. Band shading is luminance-matched across panels so that equal dispersion renders at equal density. Forecastability is heterogeneous across both horizon and semantic domain: the profiles differ in level and in decay rate, dispersion is wide within every category, and all six panels show a local seasonal peak near h = 12.}\label{fig:amiprofiles}
\end{figure}

We first examine whether AMI profiles differ across M4 Monthly categories. Fig.~\ref{fig:amiprofiles} reports median AMI by horizon and category. AMI profiles vary systematically across categories: at h = 1, median AMI ranges from 1.27 nats (Industry) to 1.61 nats (Other), with Micro (1.29), Macro (1.30), Demographic (1.31), and Finance (1.40) occupying intermediate positions. Across all categories, median AMI declines with horizon, consistent with diminishing past-future dependence as the forecast distance extends. The decay is steepest for Micro, whose median falls 61\% from h = 1 to h = 18, and shallowest for Other at 51\%, suggesting that forecastability is more concentrated at short horizons for Micro series than for the remaining categories.

These category-level AMI differences motivate the stratified error analysis below.

\subsection*{AMI and realised MASE performance}

Table~\ref{tab:skill} reports the mean Spearman rank correlation between AMI estimated from training data and the negation of realised out-of-sample error across all 18 horizons, for each model. Positive values confirm that higher AMI is associated with lower realised error.

\begin{table}[t]
\caption{Association between AMI and realised MASE skill (mean Spearman $\rho$ across horizons, with series-level bootstrap 95\% intervals).}\label{tab:skill}
\begin{tabular*}{\textwidth}{@{\extracolsep{\fill}}lrr@{}}
\toprule
Model & Mean Spearman $\rho$ & Bootstrap 95\% CI \\
\midrule
Seasonal na\"ive & $-0.1038$ & $[-0.1097,\,-0.0976]$ \\
ETS & $0.1121$ & $[0.1065,\,0.1183]$ \\
N-BEATS & $0.1259$ & $[0.1197,\,0.1321]$ \\
\bottomrule
\end{tabular*}
\end{table}

Under MASE, the AMI-skill association is positive for ETS (0.1121) and N-BEATS (0.1259), with N-BEATS showing the strongest association of the three models. Seasonal naïve shows a negative MASE correlation (-0.1038); this arises because the MASE denominator is itself an in-sample seasonal naïve error, so for high-AMI series with strong seasonality the small training-period seasonal naïve error shrinks the denominator and amplifies realised MASE even when absolute errors are low. This is a known property of MASE applied to a seasonal benchmark model, and is confirmed by the naïve-1 control reported in Methods, whose numerator is not tied to the seasonal naïve scaling denominator and which shows a positive AMI-skill correlation with no sign inversion. It does not undermine the diagnostic value of AMI for the statistical and machine-learning models, which are the relevant candidates when modelling effort is being allocated. A within-horizon permutation surrogate null (1{,}000 permutations; Methods) places all observed correlations far outside the null band, whose widest bound across the six pairs is $\pm 0.0022$ (add-one permutation $p = 0.001$ for every model-diagnostic pair, the minimum attainable with 1{,}000 permutations, below the Bonferroni-adjusted threshold $\alpha = 0.0083$ for six two-sided tests). These correlations are read for direction and consistency, not magnitude: $\rho$ of 0.11 to 0.13 corresponds to between 1 and 2\% of rank variance, and at $n = 47{,}992$ a permutation test can only exclude an exactly zero association, so statistical significance here carries no evidence of practical magnitude. The decile gradient reported below is the descriptive aggregation of this weak-but-systematic association, and the horizon-wise, category-level, and decile analyses that follow are likewise reported as descriptive, without further multiplicity adjustment.

Training length is a potential common cause of this association, since series length shifts both the KSG estimate and, in principle, realised accuracy. A dedicated control analysis (Supplementary Note~2 and Supplementary Table~S2) finds the opposite of a confound. Training length correlates with both diagnostics (Spearman 0.4574 with AMI and 0.4705 with absolute autocorrelation) but is essentially uncorrelated with realised ETS and N-BEATS skill ($|\rho| < 0.01$), so the pathway required for confounding is absent. Partial correlations controlling the rank of training length do not attenuate the association (ETS 0.1121 to 0.1300, 95\% CI [0.1246, 0.1361]; N-BEATS 0.1259 to 0.1417, 95\% CI [0.1356, 0.1475]); the modest increase is largely the mechanical variance deflation that follows from partialling out a variable correlated with the diagnostic but not the outcome, and is not read as evidence of a stronger effect. The association is positive within every training-length quartile (Supplementary Table~S3). Length therefore acts as diagnostic-side noise rather than as a confounder.

\subsection*{The AMI decile gradient}

The decile analysis converts the continuous AMI diagnostic into a stratified description of realised error. For each horizon, series are assigned to within-horizon AMI deciles (1 = lowest AMI, 10 = highest; equal-count bins on the within-horizon ranks), and median MASE is computed per decile, model, and horizon. A useful forecastability measure should produce a graded decline in realised error across deciles for models capable of exploiting temporal structure.

\begin{figure}[tp]
\centering
\includegraphics[width=\textwidth]{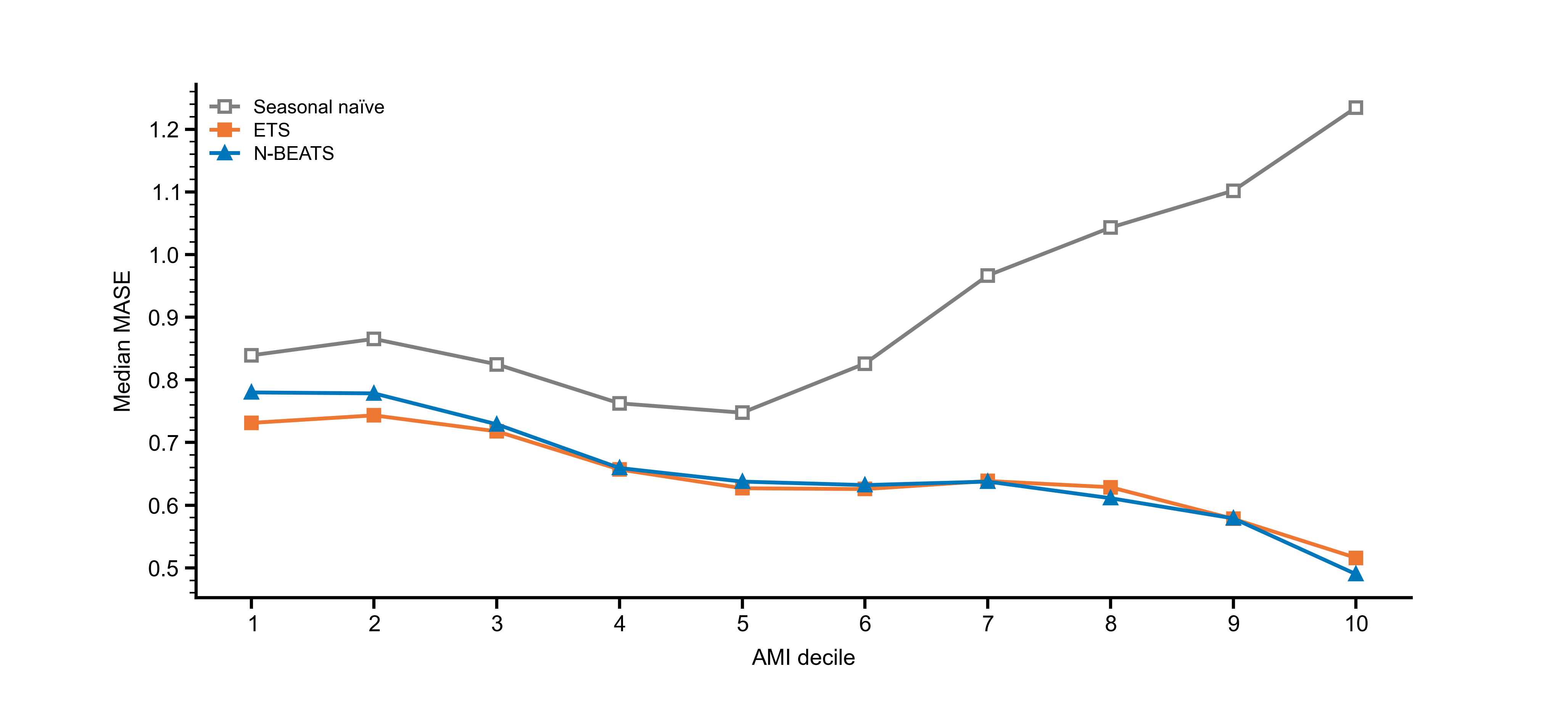}
\caption{Median MASE by AMI decile. Mean of per-horizon median MASE by within-horizon AMI decile (1 = lowest AMI, 10 = highest) for seasonal naïve, ETS, and N-BEATS over horizons 1 to 18. Median MASE declines with AMI decile for ETS and N-BEATS (a graded rather than strictly monotone decline; see text) and rises for seasonal naïve, consistent with the MASE-denominator effect.}\label{fig:decilegradient}
\end{figure}

Fig.~\ref{fig:decilegradient} reports the decile gradient. Averaged over horizons, median MASE declines from 0.731 (decile 1) to 0.516 (decile 10) for ETS and from 0.780 to 0.490 for N-BEATS, reductions of 29\% and 37\%, with rank correlations between decile index and median MASE of $-0.891$ and $-0.964$ respectively. The headline decile profile, defined at each decile rank as the median across the 18 horizons of the per-horizon decile medians, gives the same picture (ETS 0.741 to 0.531; N-BEATS 0.801 to 0.494). Seasonal naïve inverts (0.839 rising to 1.235; rank correlation $+0.661$), consistent with the MASE-denominator effect noted above. The decline is graded rather than strictly monotone: the headline profile is not strictly monotone for any probe, with a local rise around deciles 6 to 7 for the statistical and machine-learning probes, and a strictly monotone profile holds at 17\% of individual horizons for ETS and 22\% for N-BEATS (0\% for seasonal naïve). Two robustness properties support the gradient reading. The graded decline across ten bins does not depend on any particular binning choice, reflecting a continuous underlying association between the diagnostic and realised error; and a pooled-pair construction, binning all series-horizon pairs into a single equal-count decile ladder, agrees in direction with the within-horizon construction for every probe.

Supplementary Fig.~S1 shows the decile gradient within each of the six M4 categories; the within-category pattern mirrors the pooled result, indicating that the association is not a proxy for category membership, with full category-by-horizon tables in the archived tables.

The series-level correlations and the decile gradient should be read together, because their magnitudes differ for a structural reason. Realised MASE for a single series-horizon pair has a single-observation numerator, so the series-level outcome is extremely noisy, and outcome noise of this kind attenuates any series-level rank correlation towards zero; associations of the size reported in Table~\ref{tab:skill} are what a systematic but heavily attenuated relationship looks like at this granularity. The decile analysis aggregates that noise away: group medians expose the underlying gradient, which is why modest series-level correlations coexist with top-to-bottom decile reductions in median MASE of 29\% and 37\%. The series-level correlations remain the primary inferential quantity; the decile medians are the descriptive quantity, exposing the size of the error separation that the correlations summarise.

\subsection*{Comparison with absolute autocorrelation}

Table~\ref{tab:amiacf} reports the mean Spearman correlation with MASE skill for AMI and absolute autocorrelation side by side, together with the difference and its paired bootstrap 95\% interval (series-level resampling on shared resamples, B = 10{,}000; Methods), for each model. All three intervals exclude zero.

\begin{table}[t]
\caption{AMI versus absolute autocorrelation as horizon-specific diagnostics (mean Spearman $\rho$ across horizons). Differences are computed before rounding.}\label{tab:amiacf}
\begin{tabular*}{\textwidth}{@{\extracolsep{\fill}}lrrrr@{}}
\toprule
Model & Mean $\rho$ (AMI) & Mean $\rho$ ($|\mathrm{ACF}|$) & Difference & 95\% CI on difference \\
\midrule
Seasonal na\"ive & $-0.1038$ & $-0.1200$ & $+0.0162$ & $[+0.0139,\,+0.0185]$ \\
ETS & $0.1121$ & $0.0928$ & $+0.0193$ & $[+0.0169,\,+0.0216]$ \\
N-BEATS & $0.1259$ & $0.1115$ & $+0.0144$ & $[+0.0121,\,+0.0168]$ \\
\bottomrule
\end{tabular*}
\end{table}

Under MASE, AMI outperforms absolute autocorrelation, by a small but consistent margin, for all three models across the full 18-horizon range (ETS: +0.0193; N-BEATS: +0.0144; seasonal naïve: +0.0162). These differences are small relative to the headline associations and should be read as a consistent directional advantage of modest size. The advantage emerges from h = 4 onwards for ETS and seasonal naïve, and h = 5 for N-BEATS, and is most pronounced at longer horizons where absolute autocorrelation becomes near-zero or slightly negative whilst AMI retains a positive association. This horizon pattern is not metric-specific: under RMSSE the advantage is likewise concentrated at longer horizons (Supplementary Fig.~S2). It is also robust to the cells where AMI estimation is least reliable: the advantage persists in sign for every probe when horizons 13 to 18 are excluded, and strengthens when the shortest quarter of series is dropped (Supplementary Table~S7). At horizons 1 to 12 alone the advantage is smaller but positive for every probe (ETS $+0.0130$, N-BEATS $+0.0091$, seasonal naïve $+0.0120$), so the long horizons enlarge the advantage rather than supplying it; the small-sample validation reported below shows the estimator retains rank fidelity at exactly those effective sample sizes. At category level the evidence is corroborating for ETS and directional for N-BEATS: the advantage is positive in all twelve category-model cells (six M4 categories by ETS and N-BEATS), but the paired bootstrap interval excludes zero in five of the six categories for ETS and four (Demographic, Finance, Industry, and Micro) for N-BEATS, with the Other category resting on 269 series for ETS and 266 for N-BEATS (per-category values in the archived tables). Nor is the advantage an artefact of differential length sensitivity: absolute autocorrelation is the more length-loaded of the two diagnostics, and the length-adjusted advantage exceeds the unadjusted one (ETS $+0.0245$, 95\% CI [0.0218, 0.0270]; N-BEATS $+0.0186$, 95\% CI [0.0160, 0.0210]; Supplementary Table~S2). MASE's seasonal normalisation removes the scale effects that absolute autocorrelation partly captures and isolates the temporal-dependence structure that AMI measures directly.

\begin{figure}[tp]
\centering
\includegraphics[width=\textwidth]{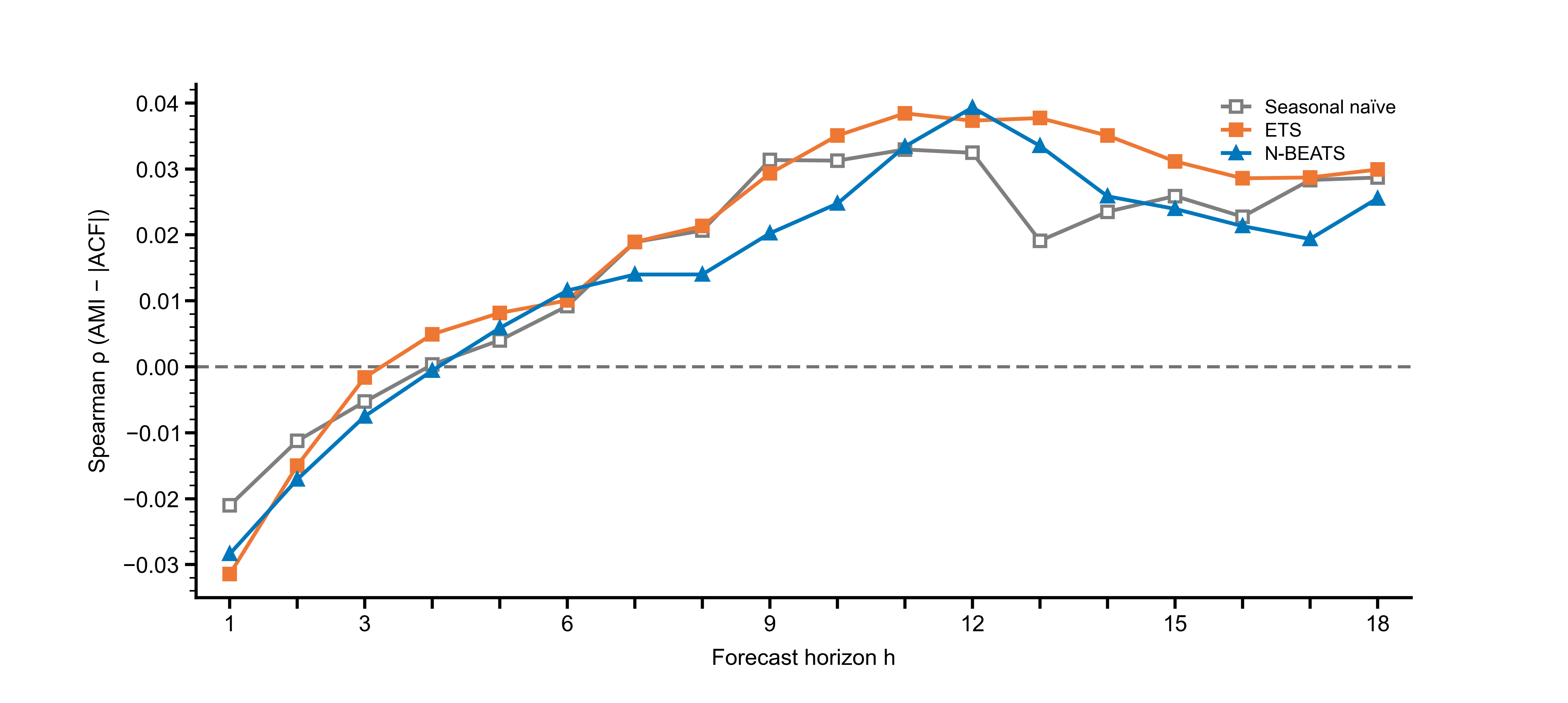}
\caption{Rank-association advantage: AMI versus absolute autocorrelation by forecast horizon. Spearman correlation difference for each model and horizon under MASE skill. Positive values indicate AMI is the stronger diagnostic at that horizon.}\label{fig:advantage}
\end{figure}

Category-level and series-level evidence point the same way: at h = 1, absolute autocorrelation is more uniform across M4 Monthly categories than AMI, ranging from 0.953 (Industry) to 0.971 (Micro), whilst AMI ranges from 1.27 to 1.61 nats. The additional category-level variation captured by AMI is not visible in the linear comparator.

Since temporal dependence typically weakens with lag, information-theoretic treatments of forecastability \cite{delsole2004,catt2009} suggest that forecastability signals may plausibly be stronger at short horizons. As supplementary evidence, the horizon-wise Spearman correlations in Fig.~\ref{fig:advantage} confirm that the AMI-skill association is positive across the 18 horizons for ETS and N-BEATS under MASE, with the AMI-over-autocorrelation advantage concentrated at longer horizons, as shown above.

\subsection*{Estimator validation on synthetic benchmarks}

The KSG estimator's accuracy at M4-typical series lengths is validated on synthetic benchmarks whose true AMI is analytically exact by construction (Fig.~\ref{fig:synthetic}; the full design, data-generating processes, and seeds are given in Supplementary Table~S5). Across 108 exact-truth cells at T = 240, the median-$k$ estimate recovers true AMI with mean absolute error 0.027 nats (bias +0.012, RMSE 0.040) and a Spearman correlation with truth of 0.93 across non-zero cells (0.87 across all cells); at T = 120 the mean absolute error is 0.041 nats with Spearman 0.90 across non-zero cells (0.87 across all cells). On white noise the median estimate never exceeds 0.006 nats after truncation, providing an empirical floor consistent with the zero-truncation rule of the estimator specification. The estimator recovers the seasonal AR(12) peak at h = 12 (0.506 against a true value of 0.511) and tracks monotone transforms with a mean absolute deviation of 0.041 nats, concentrated in the highest-AMI cells (Fig.~\ref{fig:synthetic}d); the underlying population quantity is exactly invariant, so the deviation is estimator error, not a property of AMI. Because the M4 panel admits series down to T = 48, a small-sample extension reruns the exact-truth grid at series lengths 48, 68, and 118, spanning 30 to 100 lag pairs at h = 18, the range where the panel's long-horizon estimates live and zero-truncation concentrates. Accuracy degrades gracefully (mean absolute error 0.065 nats at 30 pairs against 0.027 at T = 240) whilst rank recovery, the property the analysis relies on, remains high: the Spearman correlation with truth over non-zero cells is 0.87 to 0.93 overall and 0.92 to 0.96 at h = 13 to 18 (Supplementary Note~3, Supplementary Fig.~S3, and Supplementary Table~S8). The white-noise floor is also reported before truncation: raw estimates on white noise are centred on zero at every length (mean $-0.0002$ nats at T = 48 and $+0.0010$ at T = 240, with 50 to 57\% of raw estimates negative), so the near-zero truncated floor reflects an unbiased estimator rather than a property manufactured by the truncation rule. A tracking comparison against truth shows AMI matching absolute autocorrelation on Gaussian linear processes (0.75 against 0.74; paired difference $+0.007$, 95\% CI $[-0.027,\,+0.039]$) and exceeding it on monotone-transformed processes (0.67 against 0.52; paired difference $+0.155$, 95\% CI $[+0.105,\,+0.203]$; series-clustered bootstrap, B = 10{,}000). This supplies a mechanism for the M4 advantage reported earlier: AMI's gain over the linear diagnostic is statistically indistinguishable from zero where the dependence is linear and concentrates where the dependence structure is nonlinear.

\begin{figure}[tp]
\centering
\includegraphics[width=\textwidth]{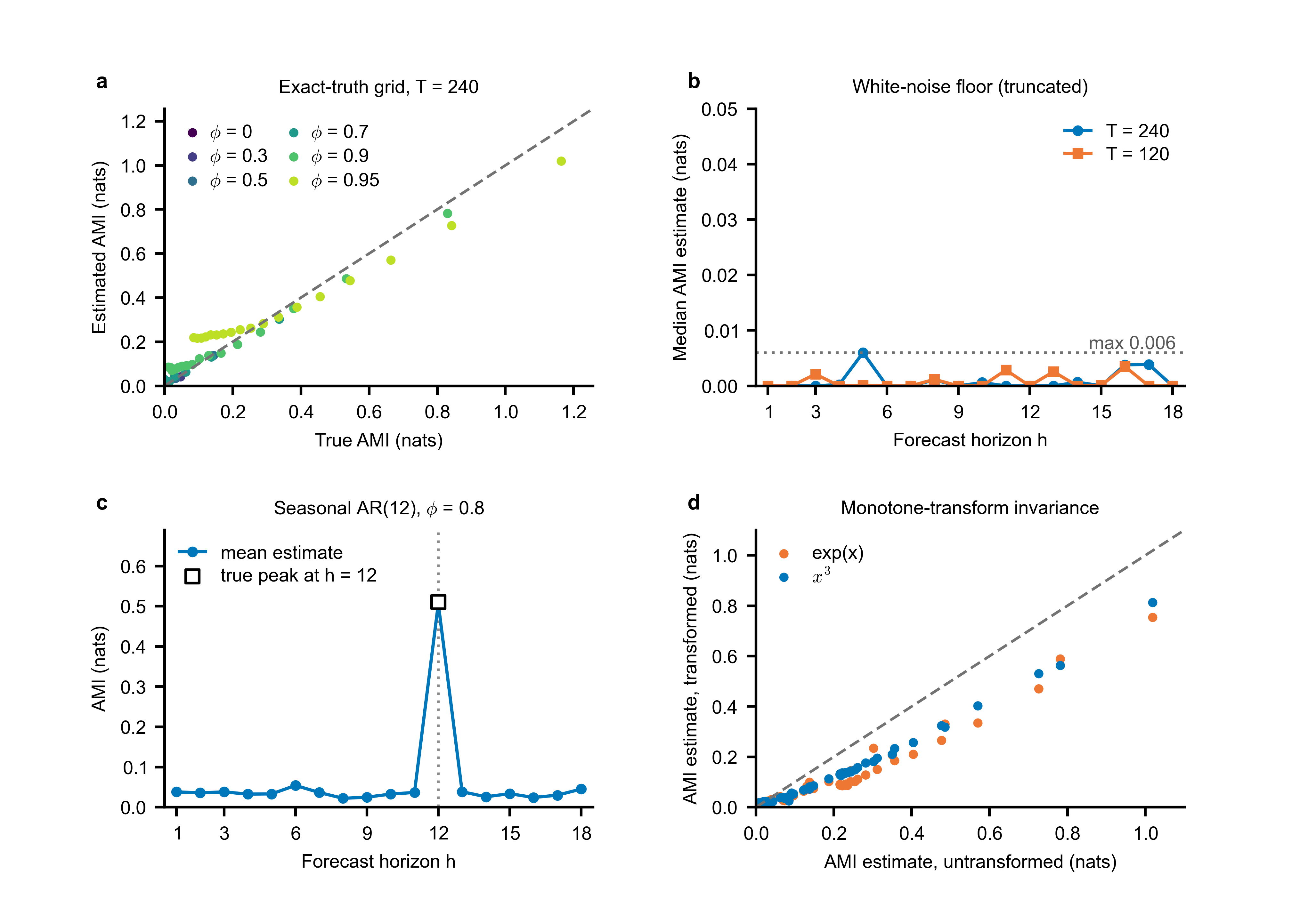}
\caption{Estimator validation on synthetic benchmarks with exact true AMI. (a) Recovery of true AMI by the median-$k$ KSG estimate across the exact-truth grid at T = 240; (b) the white-noise floor after truncation; (c) seasonal AR(12) peak recovery at h = 12; (d) monotone-transform invariance. Full design and seeds are given in Supplementary Table~S5.}\label{fig:synthetic}
\end{figure}

\section*{Discussion}

The payoff is a measurement made before modelling. AMI provides a pre-modelling, horizon-specific reading of how much recoverable temporal structure a series carries, available across a heterogeneous portfolio before any realised accuracy is observed. Under MASE, AMI provides the stronger diagnostic signal of the two compared here, modestly but consistently outperforming absolute autocorrelation across models and horizons once scale effects are removed by seasonal normalisation. A methodological point also follows for evaluation studies more generally: diagnostic-skill correlations for the model that supplies the MASE scaling denominator are structurally distorted by the shared denominator, as shown by the seasonal naïve inversion in the Results and its absence for naïve-1 in the Methods, so the scaling model should not be used as a probe when validating pre-modelling diagnostics under MASE.

Underlying this is a temporal asymmetry. Accuracy is observed only after forecasting, whereas forecastability can be assessed before it. That asymmetry is the core advantage of AMI over any retrospective model comparison: when a new series enters a portfolio, its AMI profile can be read before any model is trained.

What the measure means follows from the forecastability-competence distinction. Low AMI does not mean a series should not be forecast; it means the measured single-lag dependence is weak at the relevant horizon, so any forecasting skill must come from information beyond that signal, such as the remainder of the series history, structure visible only in combinations of lags, pooled estimation across related series, external predictors, hierarchical structure, or judgemental input. High AMI, conversely, does not guarantee low error: temporal dependence is detectable and the record holds more information to exploit, but model competence still matters, and a high-AMI series handled by an inappropriate model may still err badly. What high AMI signals is a more favourable environment for model-based forecasting.

Reporting the measure horizon by horizon, rather than as a single scalar, is deliberate. Operational series are typically non-stationary in level, variance, or dependence, and past-future coupling generally differs across horizons. Horizon-resolved AMI distinguishes series whose information decays quickly from those whose seasonal coupling re-emerges at multi-step horizons, and identifies the horizons at which a series is most forecastable. Because AMI is estimated non-parametrically from the empirical joint distribution of the paired present and future values, the estimator assumes neither linearity, Gaussianity, nor strict stationarity; it requires only that this joint distribution be informative on the training window.

The measure is computed once from training data without iterative fitting, cross-validation, or hold-out evaluation, using parallel nearest-neighbour estimation across the panel, so it scales to portfolios of tens of thousands of series. One implication lies outside the present claims: a pre-modelling forecastability measure of this kind could inform where additional modelling effort is likely to encounter exploitable temporal structure, and developing and evaluating such decision layers is future work.

Five limitations apply. First, AMI estimation can be noisy for short series; the estimator sensitivity analysis, the truncation characterisation (concentrated at long horizons and short series; Supplementary Table~S6), the small-sample validation (Supplementary Fig.~S3), and the robustness slices excluding the affected cells (Supplementary Table~S7) show this does not materially affect the results. Second, the diagnostic sees only a single lagged observation: a low-AMI series may be genuinely forecastable from the remainder of the series history, from structure visible only in combinations of lags, from pooled estimation across related series, or from external drivers, hierarchy, or calendar effects the screen cannot capture. The reading is therefore one-sided: high AMI evidences recoverable structure, whereas low AMI does not establish its absence. Third, the study establishes association, not causation, and AMI is not a sufficient statistic for model selection. Fourth, results are for M4 Monthly and may differ for other frequencies, count, or intermittent series. Fifth, the decile assignments are sample-relative and not portable to a new portfolio without recalibration against its own series distribution; a data-independent forecastability scale remains an open problem. The model panel is bounded in the same way: N-BEATS serves as a practical representative of flexible nonlinear models, not a claim of optimality; the panel spans representational capacity without optimising accuracy, and the contribution is the diagnostic, not the model comparison. Because every probe conditions on more than the diagnostic does, and N-BEATS additionally pools across series, the reported associations are conservative: a single-lag within-series statistic ranks the realised error of models with strictly greater access.

This paper proposed AMI as a pre-modelling measure of horizon-specific forecastability and validated it at scale on M4 Monthly. Realised error declines across within-horizon AMI deciles for the statistical and machine-learning probes, and AMI outperforms absolute autocorrelation, by a small but consistent margin, under MASE, with the advantage concentrated where the dependence structure is nonlinear. The broader implication is measurement-first: before asking which model should be trained, one can ask whether the series-horizon pair contains exploitable temporal information. Future work extends the diagnostic to other frequencies and to intermittent series, combines AMI with external-predictor information where the univariate signal is weak, and monitors AMI profiles over time as an early-warning layer for structural change.

\section*{Methods}

\subsection*{The diagnostic}
\subsubsection*{Overview}

The diagnostic operates in three steps. First, the horizon-specific AMI profile is estimated for each series from training data only. Second, an absolute autocorrelation profile is estimated from the same training window as a linear benchmark. Third, forecasts are generated on the official holdout and realised error is measured under MASE, stratified by within-horizon AMI decile.

\subsubsection*{The pre-modelling question}

For each series i and horizon h in a portfolio of N series evaluated over H horizons, the question is whether the declared information set contains exploitable temporal information at that horizon. It cannot be fully resolved by post-hoc accuracy analysis because accuracy is only observable after model deployment. A pre-modelling signal that identifies series-horizon pairs with richer or poorer temporal structure therefore has direct value as a measurement, even if imperfect: it characterises the informational richness of the historical data independently of which specific model is selected.

\subsubsection*{Decile stratification}

For each horizon h, series are stratified into ten within-horizon AMI deciles, formed as equal-count bins on the within-horizon ranks of the AMI estimates; rank binning keeps the bins equal-count in the presence of ties from zero-truncated estimates at long horizons. This ordinal approach avoids overclaiming precision from the AMI point estimate and is robust to estimation noise in individual series. The decile boundaries are computed within-horizon, so the classification is relative rather than absolute. The headline decile profile reports, at each decile rank, the median across the 18 horizons of the per-horizon decile medians of MASE, per model, alongside whether that profile is strictly monotone and the fraction of individual horizons whose own profile is strictly monotone. A pooled-pair robustness construction bins all series-horizon pairs into a single equal-count decile ladder; agreement in direction between the pooled and within-horizon constructions is reported alongside the profile.

\subsubsection*{Forecastability and realised error}

For model m, series i, and horizon h, the analysis observes forecast error e(m,i,h) under MASE evaluated on the official holdout.

The analysis is strictly diagnostic. It does not estimate model gain over a benchmark or test whether any model improves on seasonal naïve. It asks whether AMI identifies forecastability structure that corresponds to systematically lower or higher realised error, across all models in the panel.

\subsubsection*{Why autocorrelation is the comparator}

Autocorrelation is the natural baseline comparator because it is the most widely used linear dependence diagnostic, it is horizon-specific in the same sense as AMI, and it is computationally trivial. Scalar diagnostics from the entropy family, including approximate, sample, permutation, and spectral entropy, assign a whole series a single number and therefore cannot be evaluated at the series-by-horizon grain on which this study operates; absolute autocorrelation is the one widely used dependence diagnostic that shares AMI's horizon index. The absolute autocorrelation at horizon h, defined in Eq.~\ref{eq:acf}, is computed from the same training window as AMI.

The empirical question is whether AMI provides a stronger or more reliable association with realised forecast error than $|\mathrm{ACF}|$, particularly at short horizons where nonlinear dependence may be the primary structure. If AMI and autocorrelation are equally informative, the simpler diagnostic suffices. If AMI adds diagnostic value, the additional estimation cost is justified.

\subsection*{Data and forecasting models}
\subsubsection*{Dataset}

The empirical analysis uses the M4 Monthly dataset \cite{makridakis2020}. The M4 competition is the largest of the M competitions and the canonical benchmark in the forecasting literature. In this paper it is used as a heterogeneous portfolio benchmark: a single collection spanning six unequal categories, with substantial variation in series length, scale, seasonal strength, and signal-to-noise ratio across business, economic, financial, demographic, and industry domains. This heterogeneity is the analytics setting the diagnostic is designed for, namely mixed and uneven time-series collections where a single modelling pipeline applied uniformly is likely to misallocate effort. The Monthly subset is selected for three reasons. First, Monthly is the most widely used forecasting frequency across applied settings: Spiliotis et al. \cite{spiliotis2020} document that it carries the highest priority rating across Finance, Industry, Macro, and Micro domains simultaneously, spanning commercial, governmental, and economic forecasting contexts. The M4 Monthly panel itself covers six semantic categories (Micro, Macro, Industry, Finance, Demographic, and Other), reflecting this breadth, and making it the most practically relevant frequency for a diagnostic evaluated across a diverse portfolio. Second, with 48,000 series spanning six semantic categories, the Monthly subset is the largest single-frequency stratum in M4, providing the panel size required to detect horizon-specific AMI--error associations reliably, to report category-stratified results with adequate coverage, and to produce bootstrap confidence intervals of meaningful width. Third, the 18-month official holdout provides 18 distinct horizon-specific evaluation points per series, which is essential for characterising how the AMI--error relationship evolves across forecast distance instead of collapsing it to a single average.  Spiliotis et al. \cite{spiliotis2020} validated M4's representativeness against earlier competition datasets via instance space analysis, confirming broad coverage of the trend, seasonality, spectral-entropy, and distributional feature space; this breadth makes it a demanding testbed, since the diagnostic must behave consistently across a wide variety of temporal structures. Forecasts are evaluated on the official M4 Monthly 18-month holdout. After the inclusion rules below, the panel comprises 47,992 series (Micro 10,975; Finance 10,987; Industry 10,017; Macro 10,016; Demographic 5,728; Other 269), giving 863,856 series-horizon observations.

\subsubsection*{Inclusion rules}

Series are included if they meet four conditions. First, the training window must be sufficient to estimate AMI at all 18 horizons under the bivariate KSG protocol, requiring a minimum effective sample after lag-pairing. Second, the series must have a valid 18-month holdout aligned with the M4 competition protocol. Third, the series must have non-degenerate values for the reported error metrics, including a non-zero seasonal naïve scaling denominator for MASE; the MASE scale is computed from the training series using mean absolute seasonal differences at period 12. Fourth, both AMI and autocorrelation must be computable without numerical failure. No subjective filters for noise level, volatility, or category are applied; all exclusions are on numerical grounds only. Of the 48,000 M4 Monthly series, 8 fail the first condition: each has fewer than 48 training observations, which is insufficient to form the minimum 30 lag pairs required by the KSG-I estimator at the maximum horizon h = 18 (since the number of lag pairs at horizon $h$ equals $T-h$, so forming the minimum 30 pairs at $h=18$ requires $T-18 \geq 30$, that is $T \geq 48$). These 8 series, all in the Other category, represent 0.02 \% of the panel; their omission has no material effect on the results.

\subsubsection*{Forecasting models}

The model panel is deliberately compact. Its purpose is to span a range of representational capacity, from a minimal seasonal benchmark through higher-capacity probe models to a nonlinear deep learning approach, not to identify the globally optimal model for M4 Monthly series.

Seasonal naïve forecasts for monthly data repeat an observed seasonal cycle. For horizon h at origin T, the forecast is the observation at position T + h - 12 for h up to 12, and at position T + h - 24 for h from 13 to 18, so that every forecast is drawn from an observed value. This is the appropriate seasonal anchor for monthly data and is hard to beat substantially in weakly forecastable series.

ETS is estimated independently for each series. The selection procedure evaluates the full admissible grid of state-space models with additive error throughout: additive and, for strictly positive series, multiplicative trend and seasonal components, with and without damping, yielding up to 15 configurations per series. The model with the lowest AICc is selected; AIC is used as a fallback when AICc is non-finite. This follows the procedure of Hyndman and Khandakar \cite{hyndman2008}; exponential smoothing is reviewed by Gardner \cite{gardner2006} and given textbook treatment by Hyndman and Athanasopoulos \cite{hyndman2021}. ETS provides a principled classical statistical reference model with explicit level, trend, and seasonal components.

N-BEATS \cite{oreshkin2020} is a deep learning forecasting model built from residual stacks of fully connected layers. The implementation uses the generic architecture, with identity stacks and no basis expansion. A single model is estimated globally on the pooled M4 Monthly training series, with parameters shared across series rather than fitted series by series as ETS is, and emits all 18 horizons jointly in a direct multi-step configuration. N-BEATS requires no hand-crafted features and learns temporal patterns directly from the raw time series, making it a strong test of whether AMI captures structure that a neural architecture can also exploit. The model was trained following the M4 competition protocol, with hyperparameters set per Oreshkin et al. \cite{oreshkin2020}; the input lookback window is 48 observations (four seasonal periods). Series with fewer than 49 training observations cannot form a training window and receive no N-BEATS forecast, so they are absent from the N-BEATS columns throughout (3 series, all in the Other category, at all 18 horizons; the count is recorded in the run log). The full training configuration (architecture, optimiser, batch size, training steps, loss, and random seed) is given in Supplementary Table~S4. N-BEATS is included as a nonlinear deep learning capacity probe, not the claimed contribution. Unlike ETS and seasonal naïve, N-BEATS attains accuracy on the M4 dataset competitive with the leading neural and hybrid approaches \cite{smyl2020}, which strengthens the diagnostic test: if AMI reliably predicts error for N-BEATS, the forecastability signal is robust to representational capacity.

The model is trained on the raw training levels. Scale heterogeneity across series is handled by neuralforecast's per-window robust normalisation, which centres each input window on its median, scales it by the window's median absolute deviation, and is inverted on the model output, so no differencing and no manual rescaling is applied to the target.

\subsection*{Estimation and empirical tests}
\subsubsection*{AMI estimation}

The k-nearest-neighbour mutual-information estimator of Kraskov et al. \cite{kraskov2004,paninski2003,verdu2019} is adopted because it is non-parametric, captures nonlinear dependence, runs at portfolio scale, and produces an interpretable per-series, per-horizon feature suitable for automated computation across a panel. For each series $i$ and horizon $h$, AMI is estimated as the mutual information between the single lagged observation $Y_{i,t}$ and $Y_{i,t+h}$, formed over all lag pairs available in the training window. The estimator is a custom KSG-style nearest-neighbour estimator following Kraskov et al. \cite{kraskov2004}, using the bivariate formulation; the primary diagnostic is the median of the estimates obtained with neighbour counts k = 3, k = 5, and k = 10. The median across neighbour counts is preferred to any single k because it damps the k-dependence of the KSG bias-variance trade-off (smaller k, lower bias and higher variance; larger k the reverse); Supplementary Table~S1 shows the headline correlations move by at most about $0.013$ across the individual settings. Each lagged pair is standardised to zero mean and unit variance before estimation, the two marginals separately. The standardisation is a numerical convenience for the shared nearest-neighbour radius, not a correction: mutual information is invariant to monotone transforms of the margins and the absolute autocorrelation is scale-invariant, so neither diagnostic is altered by it.

Equation~\ref{eq:ami} defines AMI for a stationary process. Many M4 Monthly series are non-stationary in level, and no differencing or detrending is applied before estimation. The estimand is therefore the empirical past-future dependence of the observed series over the training window, rather than the AMI of an underlying stationary process, and this is deliberate. The models in the panel forecast levels and are scored on levels, so dependence contributed by trend or evolving seasonality is exploitable structure for the forecasting task, not an artefact to be removed; a diagnostic that stripped it out would measure the forecastability of a different quantity from the one the portfolio must forecast. The absolute autocorrelation comparator is computed from the same training window and responds to the same level structure, so the comparison between the two diagnostics is conditioned on a common estimand. Linear trend in particular affects both: the absolute autocorrelation of a strongly trended series approaches one at all horizons, so trend alone cannot account for the AMI advantage over the linear comparator reported in the Results. One consequence should be named explicitly: estimated on levels, AMI partially indexes trend and level persistence as well as stationary dependence; for the levels-forecasting task evaluated here that is a property of the intended estimand, not a bias to be removed.

The KSG-I estimator formula is:

\begin{equation}\label{eq:ksg}
\widehat{\mathrm{AMI}}_h = \psi(k) + \psi(N) - \big\langle \psi(n_{x,i}+1) + \psi(n_{y,i}+1) \big\rangle
\end{equation}

where $\psi$ is the digamma function, $k$ is the number of neighbours, $N$ is the number of sample pairs, and $n_{x,i}$ and $n_{y,i}$ are the counts of points falling strictly within the maximum-norm radius defined by the $k$-th neighbour of point $i$ in the joint space. Marginal counts use strict inequality, enforced in the distance domain so that the count is exact at the neighbour radius. This is KSG Algorithm 1 \cite{kraskov2004}. The estimator is defined for continuous variables, whereas M4 levels are recorded at finite precision: a screen of the panel finds at least one exactly repeated lagged pair in 35.5\% of series-horizon cells, and in 9.1\% of cells some pair recurs at least four times, so that the $k = 3$ joint neighbour radius is exactly zero, which collapses the marginal counts of the affected points to their minimum and inflates the estimate.
Because repetition reflects recording precision rather than predictability, each series is therefore dequantised once before any lagged pair is formed, by adding Gaussian noise at $10^{-10}$ times the mean absolute level, seeded from the run seed and the series identifier so the result is reproducible. This follows Kraskov et al. \cite{kraskov2004} and matches the rule applied by scikit-learn's mutual information estimators. Absolute autocorrelation is computed on the raw series, since Pearson correlation is unaffected by exact repeats. KSG Algorithm 2 uses $\psi(n_x)$ and $\psi(n_y)$ without the $+1$ correction and adds a term $-1/k$; it is a distinct estimator and must not be confused with the present implementation. KSG-I is preferred here for two reasons. First, the $+1$ correction in $\psi(n_x+1)$ and $\psi(n_y+1)$ prevents the digamma function from being evaluated at zero when marginal counts are sparse, which occurs for short series at long horizons; KSG-II's omission of the $+1$ can produce undefined or severely distorted estimates in precisely those conditions. Second, AMI is used here as an ordinal ranking signal rather than an absolute information-theoretic quantity; KSG-I's known small positive bias is approximately monotone across series and does not affect the validity of rank-based conclusions. Where the estimator produces small negative values due to finite-sample noise, these are set to zero; across the M4 Monthly panel such truncations occur predominantly at long horizons ($h \geq 13$) and for short series near the 30-pair minimum (3.4\% of finite estimates overall; frequencies by training-length band and horizon in Supplementary Table~S6), and the headline associations persist when the affected cells are excluded (Supplementary Table~S7).

AMI is estimated once per series and horizon from the base training window, which excludes the official 18-month holdout. It is not updated as the evaluation proceeds, preserving the strictly pre-modelling character of the diagnostic and ensuring out-of-sample validity: the forecastability signal is computed exclusively from data available before the forecast origin.

Sensitivity to the neighbour count is reported in Supplementary Table~S1 using the individual settings k = 3, k = 5, and k = 10. The full estimator implementation is provided in the archived code (Data availability). The M4 Monthly pipeline uses a single global random seed (42), and the synthetic validation studies use their own master seeds (20260714 for the main study and 20260720 for the small-sample extension); the run manifest in the archived code records the exact software versions (Python 3.11.14, NumPy 2.4.4, pandas 2.3.3, SciPy 1.15.3, scikit-learn 1.7.2, statsmodels 0.14.6, joblib 1.5.3, PyTorch 2.9.1, neuralforecast 3.1.2) and hardware.

\subsubsection*{Autocorrelation comparator}

The absolute sample autocorrelation at horizon h is computed from the same training window as AMI. For each series i:

\begin{equation}\label{eq:acf}
\left|\mathrm{ACF}_{i,h}\right| = \left|\operatorname{Corr}\!\left(Y_{i,t},\, Y_{i,t+h}\right)\right|
\end{equation}

where the correlation is Pearson's product-moment correlation over the available lag-paired observations. As with AMI, the autocorrelation is computed once from the training window and held fixed.

AMI is benchmarked against absolute autocorrelation to test whether a nonlinear dependence diagnostic adds value beyond a simple linear horizon-specific feature.

\subsubsection*{Error metrics}

Forecasts are evaluated on the official M4 Monthly 18-month holdout. Forecast performance is evaluated solely with MASE \cite{hyndman2006,gneiting2007}, a scale-normalised error measure. For M4 Monthly series with seasonal period m = 12:

\begin{equation}\label{eq:mase}
\mathrm{MASE}_{i,h} = \frac{\left|Y_{i,T+h} - \hat{Y}_{i,h}\right|}{\dfrac{1}{T-12}\displaystyle\sum_{t=13}^{T}\left|Y_{i,t} - Y_{i,t-12}\right|}
\end{equation}

The MASE denominator is the mean absolute seasonal difference of the training series at period 12, computed from training data only: $\tfrac{1}{T-12}\sum_{t=13}^{T} |Y_{i,t} - Y_{i,t-12}|$. MASE expresses forecast error relative to a naïve in-sample benchmark, which makes it directly aligned with comparison across series of different magnitudes, volatility levels, and category characteristics.

MASE is used as the sole forecast-error outcome because the study evaluates forecastability across a heterogeneous portfolio of monthly series. A scale-normalised error measure is required to compare diagnostic value across series with different magnitudes, volatility levels, and category characteristics. MASE expresses forecast error relative to a naïve in-sample benchmark, making it directly aligned with the question addressed here: whether a series-horizon pair contains recoverable temporal information beyond what a simple default captures. The symmetric mean absolute percentage error (sMAPE) used alongside MASE in the M4 competition \cite{makridakis2020} is deliberately not adopted: despite its name it treats equal-sized over- and under-forecasts asymmetrically, and it does not elicit any standard functional of the predictive distribution, so minimising it can reward degenerate forecasts rather than accurate ones, and recent analysis concludes that it is on these grounds either unnecessary, where a scaled measure such as the RMSE or MASE would serve, or actively misleading \cite{kolassa2026,hyndman2006}. The objective is not to reproduce the M4 competition ranking framework, but to use the M4 Monthly panel as a large heterogeneous portfolio for validating a pre-modelling diagnostic.

Seasonal naïve appears in the study in two distinct roles. First, it provides the in-sample scaling denominator for MASE, computed from in-sample seasonal differences at period s = 12, used only to normalise errors so that they can be compared across heterogeneous monthly series. Second, it is included as a forecasting probe because it represents the lowest-complexity operational forecasting method available to a forecasting portfolio. These roles are not circular: the denominator is estimated from in-sample seasonal differences, whereas all forecast errors in the numerator are computed on the out-of-sample horizon at h = 1 to 18. The seasonal naïve model's MASE therefore quantifies how its realised holdout errors compare with the series' own historical seasonal naïve error scale, rather than dividing a forecast by itself.

\subsubsection*{Empirical tests}

Three tests are applied. First, the AMI--error association. For each model m and horizon h, the Spearman rank correlation is computed between AMI estimated from training data and the negation of realised error on the holdout:

\begin{equation}\label{eq:spearman}
\rho^{\mathrm{AMI}}_{h,m} = \operatorname{Corr}_S\!\left(\mathrm{AMI}_{i,h},\, -\,e^{m}_{i,h}\right)
\end{equation}

To keep interpretation consistent, all reported correlations are between each diagnostic and forecast skill, defined as negative MASE. Positive correlations therefore indicate that higher diagnostic values are associated with lower realised forecast error.

Second, the AMI versus $|\mathrm{ACF}|$ comparison. For each model and horizon, the difference in rank association is:

\begin{equation}\label{eq:advantage}
\Delta\rho_{h,m} = \rho^{\mathrm{AMI}}_{h,m} - \rho^{|\mathrm{ACF}|}_{h,m}
\end{equation}

Positive values indicate AMI is the stronger diagnostic at that horizon. Horizon-wise differences are reported in Fig.~\ref{fig:advantage} and summarised in Table~\ref{tab:amiacf}.

Median realised error is then reported by within-horizon AMI decile, model, and M4 category, under MASE.

Bootstrap 95\% intervals for the horizon-averaged correlations use the percentile method with series-level resampling and 1{,}000 replications; the resulting intervals are reported in Table~\ref{tab:skill}. The AMI-versus-absolute-autocorrelation advantage additionally receives a paired bootstrap interval: series are resampled with replacement (B = 10{,}000, fixed seed 20260719), both horizon-averaged Spearman correlations are recomputed on each shared resample under the same aggregation, and the 95\% percentile interval of their difference is reported in Table~\ref{tab:amiacf}. The same construction is applied within each M4 category and, for the synthetic study, within each data-generating-process family with series as the resampling clusters. This paired bootstrap supplements the permutation null described below; it replaces nothing. Sensitivity of the headline correlations to the neighbour count (k = 3, 5, and 10 individually, and without zero-truncation) is reported in Supplementary Table~S1 and Supplementary Note~1; the sign and model ordering are preserved in every specification, with a largest deviation of about $0.013$.

\subsubsection*{Metric choice and denominator handling}

MASE avoids the per-forecast denominator instability of percentage errors, although it requires a non-zero in-sample seasonal naïve scaling denominator, which is handled in the inclusion rules. The AMI-skill association, the AMI-versus-absolute-autocorrelation comparison, and the decile-gradient analysis are all reported under MASE in the body of the paper. Two supplementary checks support the metric choice. First, a naïve-1 (random walk) benchmark, whose MASE numerator is not tied to the seasonal naïve scaling denominator, shows a positive horizon-averaged AMI-skill correlation ($+0.077$) with no sign inversion, confirming that the seasonal naïve inversion reported in the Results is a shared-denominator artefact and not a property of simple benchmarks. Second, under RMSSE, a second scale-normalised error measure, the sign pattern and model ordering are preserved and the AMI-over-absolute-autocorrelation advantage is preserved and larger (ETS $+0.028$, N-BEATS $+0.022$, seasonal naïve $+0.024$). At the series-by-horizon grain evaluated here RMSSE reduces to the same absolute error divided by the in-sample root mean squared seasonal difference in place of the mean absolute seasonal difference, so this check tests sensitivity to the scaling denominator rather than to the choice between absolute and squared loss. The horizon profiles are consistent across the two metrics: the AMI-skill association is front-loaded at short horizons under both MASE and RMSSE, and the AMI-over-absolute-autocorrelation advantage concentrates at longer horizons under both (Supplementary Fig.~S2), so the horizon pattern reported in the Results is not metric-specific; per-horizon values under both metrics are included in the archived tables. The AMI-versus-absolute-autocorrelation comparison is therefore conditioned throughout on scale-normalised error measurement.

\subsubsection*{Permutation surrogate null}

A within-horizon permutation surrogate null distinguishes a small association from a spurious one. For each model-diagnostic pair, the diagnostic is permuted against realised skill within each horizon and the horizon-averaged Spearman correlation recomputed; 1{,}000 permutations form the null band, with a two-sided add-one permutation p-value (minimum attainable $p = 0.001$). Every observed correlation falls outside the null band at $p = 0.001$, below the Bonferroni-adjusted threshold $\alpha = 0.0083$ for the six tests. The null is deliberately cross-sectional in scope: it tests whether the association between the diagnostic and realised skill across series could arise from chance pairing alone, which is precisely the claim it supports in the Results. It is not a within-series test of nonlinear structure against a linear-Gaussian null, as phase-randomised surrogates would provide, and no such claim is attached to it; the burden of demonstrating value beyond linear dependence is carried by the direct comparison with absolute autocorrelation and by the synthetic benchmarks, in which the AMI advantage concentrates on nonlinearly transformed processes.

\section*{Data availability}

The M4 competition dataset is publicly available from the M4 competition organisers and via the M4comp2018 R package. All result artefacts for this study, including the full analysis tables underlying every reported figure and statistic, are permanently archived on Zenodo at \url{https://doi.org/10.5281/zenodo.21752929}.

\section*{Code availability}

All analysis code is archived in the same Zenodo deposit (\url{https://doi.org/10.5281/zenodo.21752929}), with the analysis pipeline (\texttt{ami\_forecastability\_m4\_monthly.py}) as the entry point, the synthetic estimator-validation scripts, a run manifest recording exact software versions, random seeds, estimator parameters, and the full N-BEATS configuration, and a README mapping every table and figure reported here to the file and script that produces it. The deposit corresponds to the canonical run (\texttt{run\_20260801\_203622\_73477}), from which all reported results are drawn.

\section*{Author contributions}

The author was responsible for conceptualisation, methodology, software, formal analysis, investigation, data curation, writing of the original draft, review and editing, visualisation, and project administration.

\section*{Funding}

The author received no specific funding for this work.

\section*{Competing interests}

Virtual Blue Ltd, the author's affiliation, provides forecasting and analytics consultancy services; this work was conducted independently and received no client funding. The author declares no competing financial or non-financial interests in relation to the work described.

\end{document}